\documentclass[pra,twocolumn, showpacs]{revtex4}
\usepackage{graphicx}
\usepackage{dcolumn}
\usepackage{bm}
\usepackage{amsmath}

\begin{document}

\title{Tunnelling dynamics of Bose-Einstein condensate in a four wells loop shaped system}

\author{Simone \surname{De Liberato}}  
\affiliation{Laboratoire Pierre Aigrain, \'Ecole Normale Sup\'erieure, 24 rue Lhomond, 75005 Paris, France}
\author{Christopher J. Foot}  
\affiliation{Clarendon Laboratory, Parks Road, Oxford OX1 3PU, United Kingdom}

\date{\today}

\begin{abstract}
We investigated the tunnelling dynamics of a zero-temperature Bose-Einstein condensate (BEC) in configuration of four potential wells arranged in a loop.
We found three interesting dynamic regimes: (a) flows of matter with small amplitude, (b) steady flow and (c) forced flow of matter for large amplitudes. The regime of quantum self-confinement has been studied and a new variant of it has been found for this system.
\end{abstract}  

\pacs{03.75.Kk, 03.75.Lm, 74.50.+r}

\maketitle

\section{Introduction}
 
The behavior of Bose-Einstein condensate in two wells potential based on two state approximation has been investigated in a number of theoretical and experimental papers (\cite{smer97,smer99,zapa98,ragh99,jack99,mari99,sake04,albi04}). Here we 
expand this method to a four-well system with periodic boundary conditions.
The main aim of this analysis is to study possible ways in which to achieve mass transport around a loop and persistent currents.

\section{The Gross-Pitaevskii equation with the Feynman ansaz}

The behavior of a Bose-Einstein condensate at low temperature is
accurately described by a nonlinear Schr\"odinger equation, known as
the Gross-Piteavskii equation (GPE), obtained from the two bodies interaction Hamiltonian by neglecting the quantum fluctuations of the bosonic field.
 
The GPE equation that describes a BEC trapped in a potential $V(\bm{r})$ 
is:\begin{equation}
i\hbar\frac{\partial\Psi}{\partial t}=\left[-\frac{\hbar^{2}\nabla^{2}}{2m}+V(\bm{r})+g\mid\Psi\mid^{2}\right]\Psi\label{eq:GPE}\end{equation}
 
with a coupling constant $g=4\pi\hbar^{2}a/m$, where $a$ is the scattering length of atom (that has mass $m$), taken here to be positive.
 
We take the normalized solution of the time-independent GPE for the ith non interacting well to be $\Phi_{i}$ and look for an approximate solution of the form (\cite{feyn65,smer97,mahm05}):
 
\begin{equation}
\Psi(\bm{r},t)=\sum_{i=1}^{4}\psi_{i}(t)\Phi_{i}(r).\label{eq:ansaz}\end{equation}

Substituting this form into Eq.(\ref{eq:GPE}) and writing the time-dependent function $\psi_{i}$ as $\sqrt{N_{i}}e^{i\theta_{i}}$ we obtain, after an integration over the spatial variables, a system of four coupled complex
equations:

\begin{eqnarray}
\label{firstsystem}
i\hbar\partial_{t}\psi_{1}=(E_{1}^{0}+U_{1}N_{1})\psi_{1}-K_{1}\psi_{2}-K_{4}\psi_{4}\nonumber\\
i\hbar\partial_{t}\psi_{2}=(E_{2}^{0}+U_{2}N_{2})\psi_{2}-K_{2}\psi_{3}-K_{1}\psi_{1}\\
i\hbar\partial_{t}\psi_{3}=(E_{3}^{0}+U_{3}N_{3})\psi_{3}-K_{3}\psi_{4}-K_{2}\psi_{2}\nonumber\\
i\hbar\partial_{t}\psi_{4}=(E_{4}^{0}+U_{4}N_{4})\psi_{4}-K_{4}\psi_{1}-K_{3}\psi_{3}\nonumber
\end{eqnarray}

with the condition $\sum_{i=1}^{4}N_{i}=N_{T}$.
Where $N_{T}$ is the total number of particles, $E_{i}^{0}$
is the ground state energy for the ith well, $U_{i}$ is the self-interaction energy of the ith well and $K_{i}$ is a parameter that characterizes the overlap between wells. 

It is useful to recast Eq.(\ref{firstsystem}) as a system of
eight real equations:

\begin{eqnarray}
\label{realsystem}
\hbar\partial_tN_i= & -K_i\sqrt{N_iN_{i+1}}\sin(\theta_{i+1}-\theta_i)& \nonumber \\
&+K_{i-1}\sqrt{N_{i-1}N_i}\sin(\theta_i-\theta_{i-1})&  \\
\hbar\partial_t\theta_i=& 
+K_i\sqrt{N_{i+1}/N_i}\cos(\theta_{i+1}-\theta_i)&\nonumber \\
&+K_{i-1}\sqrt{N_{i-1}/N_i}\cos(\theta_i-\theta_{i-1})&-UN_i-E_i^0\nonumber 
\end{eqnarray}
 
with $i=1$ to $4$ and all the arithmetic on the index modulo $4$ ($N_5\equiv N_1$ etc.).  

As pointed out in \cite{smer97} the total number of atoms $N_{T}$
is constant but, in order to have a coherent phase description, the phase
fluctuations must be small, giving a lower bound on the number of particles, namely $N_{T}>N_{min}\simeq10^{3}$ 
\cite{gajd97}. One may notice that typically the value of $U_{i}$ is largely independent from $i$; we thus drop the index $i$ and consider all the $U_{i}$ to be the same.
In the rest of the paper we will consider only the case of all the $K_i$ modulated around the same mean value and the case of all the $K_i$ fixed to a certain positive value or set to $0$. We will call $\tilde K$ this mean or fixed value and define the two quantity $\omega_R$ and $\Lambda$, such that $\tilde K=\hbar\omega_R/2$ and $\Lambda=UN_T/2\tilde K$. The first parameter has the dimension of a frequency and we will take its inverse as time unit while the second is a dimensionless quantity regulating the behavior of the system.

\begin{figure}
\begin{center}
\includegraphics[width=2.5cm]{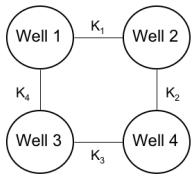} 
\caption{\label{fig1} The lay out of our system with the coupling 
constants evidenciated.}
\end{center}
\end{figure}

\section{Small amplitude regime}

\begin{figure}
\begin{center}
\includegraphics[width=4.2cm,height=3cm]{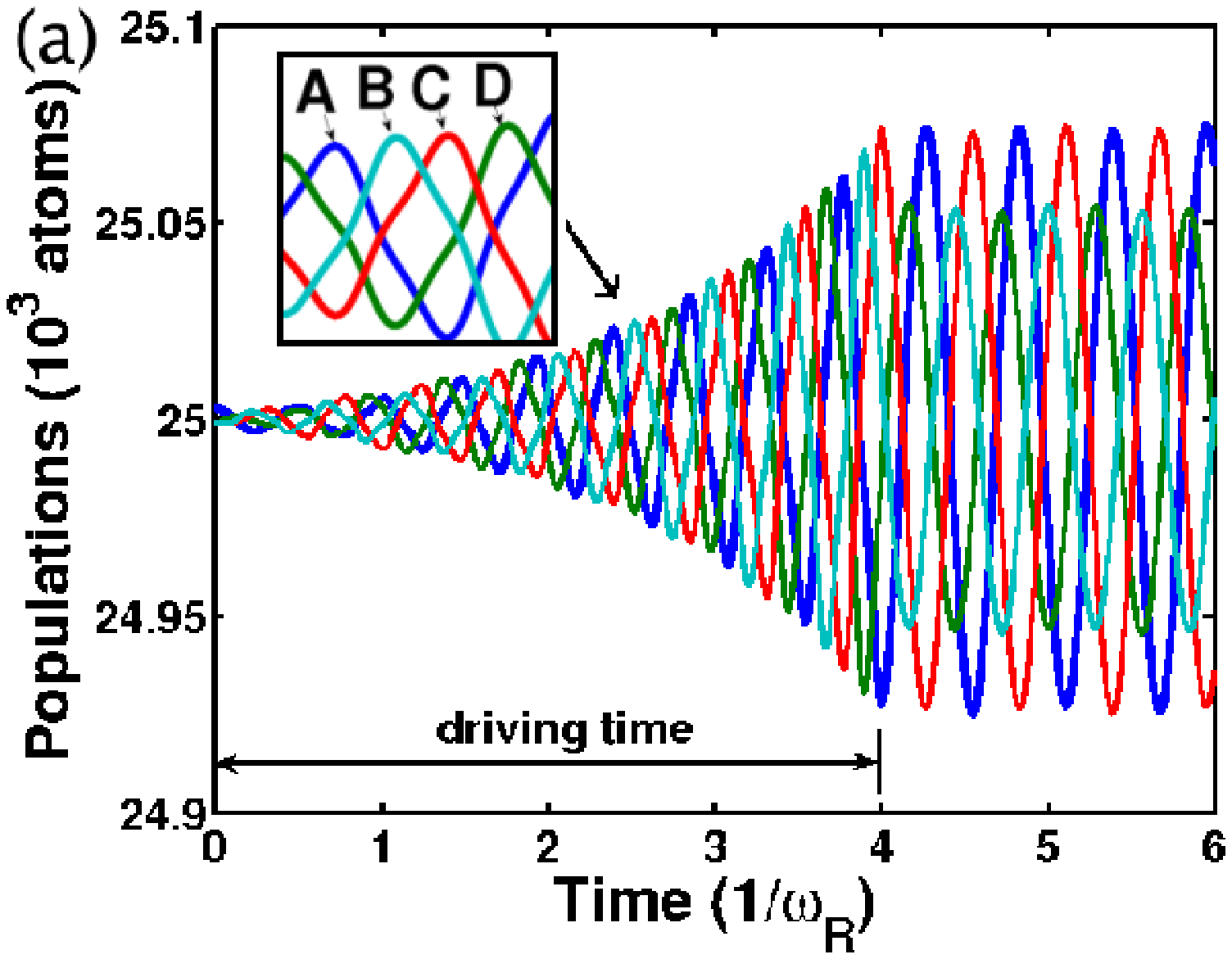}
\includegraphics[width=4.2cm,height=3cm]{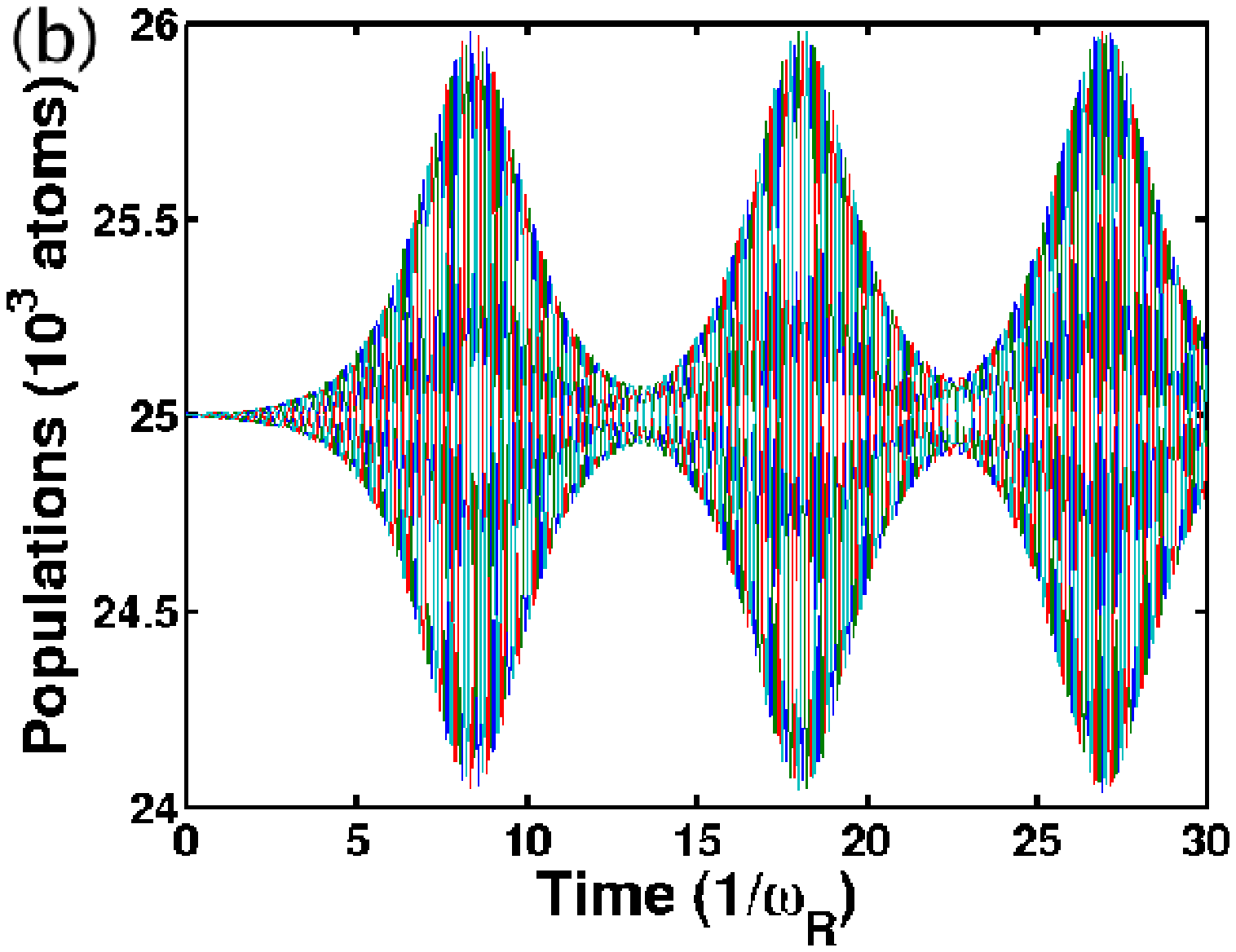}
\caption{\label{fig2}(Color online) The lines A,B,C and D are the plots of population versus time in the linear regime for the first, second, third and fourth well respectively. In (a) the mass surplus increases during the resonant driving from time $t=0$ to  $t=\tau$, where $\tau=4/\omega_R$.  
For $t>\tau$ there is a dissipationless flow around the loop. In (b) the modulation is not interrupted and the amplitude beats are visible. In these simulations $N_{T}=10^{5}$ and $\Lambda=500$.} 
\end{center}
\end{figure}

Starting with an arbitrary small population inbalance in one of the four wells we can in principle, by modulating the coupling constants, amplify it and make it spin around the ring. In the case of symmetric wells ($E_{i}^{0}=E_{j}^{0}$ for all $i,j$) we 
simulated the system with coupling constants of the form: $K_{i}=\tilde K(1+(-1)^{i}\sin(wt+\phi))$, i.e. a periodic oscillation at the frequency $w=\sqrt{3UN_{T}\tilde K+2\tilde K^{2}}$, that is the resonance frequency of the Eq.(\ref{realsystem}) linearized around 
the value $N_i=N_T/4$ and $\theta_i-\theta_{i+1}=0$ for all $i$. 
The result of our simulation is shown in Fig. \ref{fig2}; starting
with an arbitrary small population excess in one well the  resonant driving
increases the population difference and makes it spin around the loop.
We are effectively causing the system to spin by modulation of the coupling constant in the same way one could make a ball turn  in a dish by raising and lowering its edges at the right moment. The direction of the spinning is imposed by the initial phase $\phi$ of the perturbation relative to the position of the initial imbalance.
 
If we stop the modulation at $t=\tau$ we can see in  Fig. \ref{fig2} (a) that the particle surplus keeps going around the loop and will only be dumped by dissipation  (\cite{kohl03,zapa03,zapa98,sina00,kohl02}). However the dumping time is of the order of some $1/\omega_R$, substantially longer than the time-scale we are interested in. We can thus neglect it. In any case we must notice that this process can prove to be very difficult to measure in practice, due to the extremely small amplitude of the oscillations.
Indeed one can not amplify the imbalance over a certain small percentage of the average well population as in this case, due to the change of resonance frequency of the system with the amplitude of the oscillations, we see periodic beats of the amplitude ( Fig. \ref{fig3} (b)). As $\Lambda$ decreases, the peak amplitude of these beats increases but their frequency decreases. This means that, in order to have beats large enough to be observed, their period will be of the order of some tens of $1/\omega_R$ and thus the dissipation could not be a negligible problem anymore, due to the intrinsically longer time-scale implied. As stated in \cite{kohl03} the application of a periodic modulation of the potential well can stabilize the tunneling dynamics against dissipation and thus the beats in amplitude could be effectively observable. In any case, the exact study of these dissipation effects is out of the scope of our present work.

\section{Nonlinear regime}

\subsection{Phase imprinting}
In the nonlinear regime, exploiting the fact that the phase is defined modulo $2\pi$, we can have a constant dephasing between neighboring wells.
If we set all the populations at the same value $\frac{1}{4}N_T$, all the $E_i^0$ and all the $K_i$ at the same constant values and all the phase differences between adjacent wells to $\pi/2$, building up in this way a phase difference of $2\pi$ around the loop, we have that in Eq.(\ref{realsystem}) the gain and the loss term for each well equilibrates each other, and the relative phases of the wells stay constant.
Being in this case the system invariant under rotations of $\pi/2$ and
being the evolution of the populations dependent on the relative phases,
we can expect that the only possible behaviors of the system can be
a steady flow of particles around the loop or no flux at all.
In both cases, from our simulation we will see nothing but the fact that all the populations stay constant at the value of $\frac{1}{4}N_T$.

\begin{figure}
\begin{center}
\includegraphics[width=4.2cm,height=3cm]{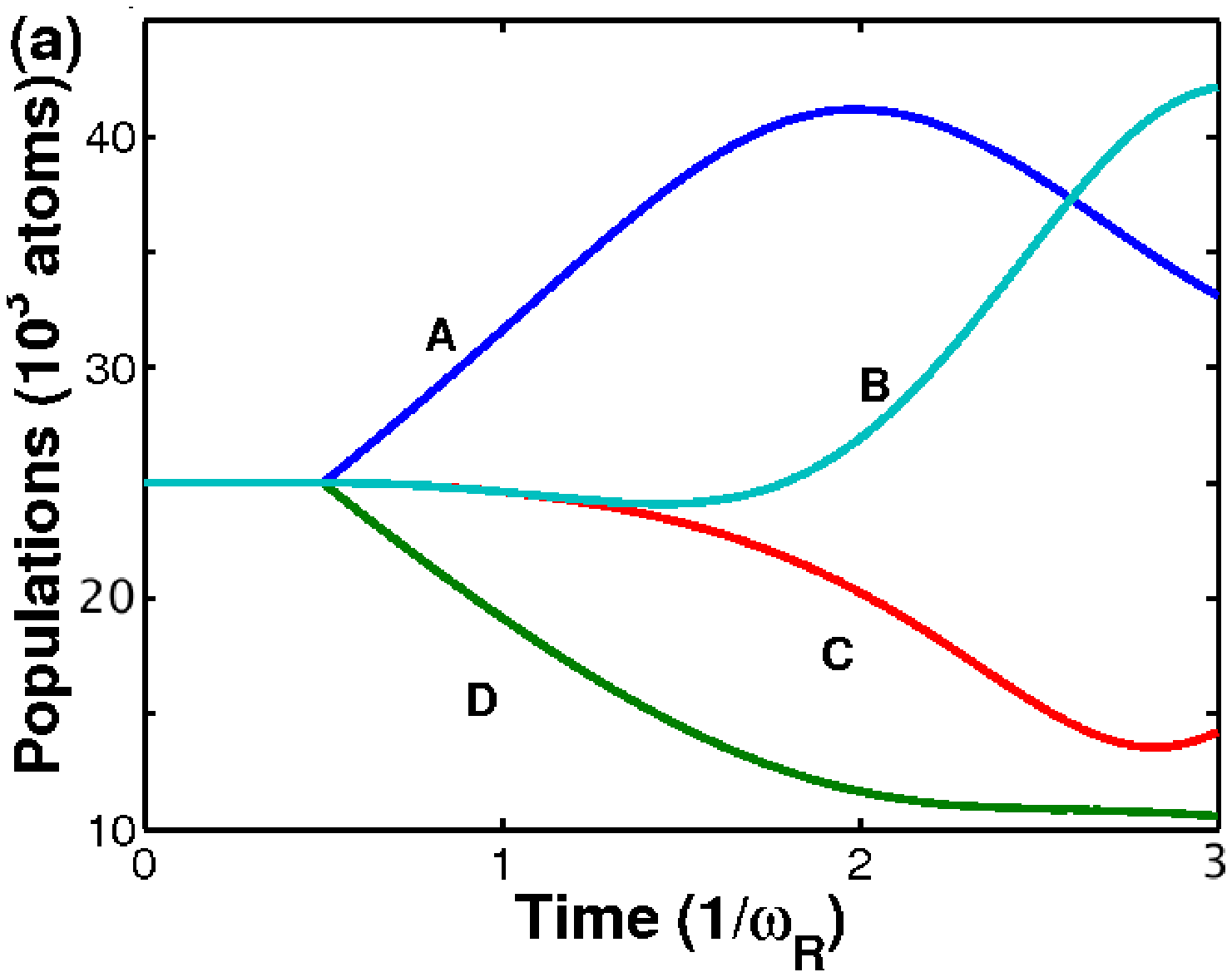}
\includegraphics[width=4.2cm,height=3cm]{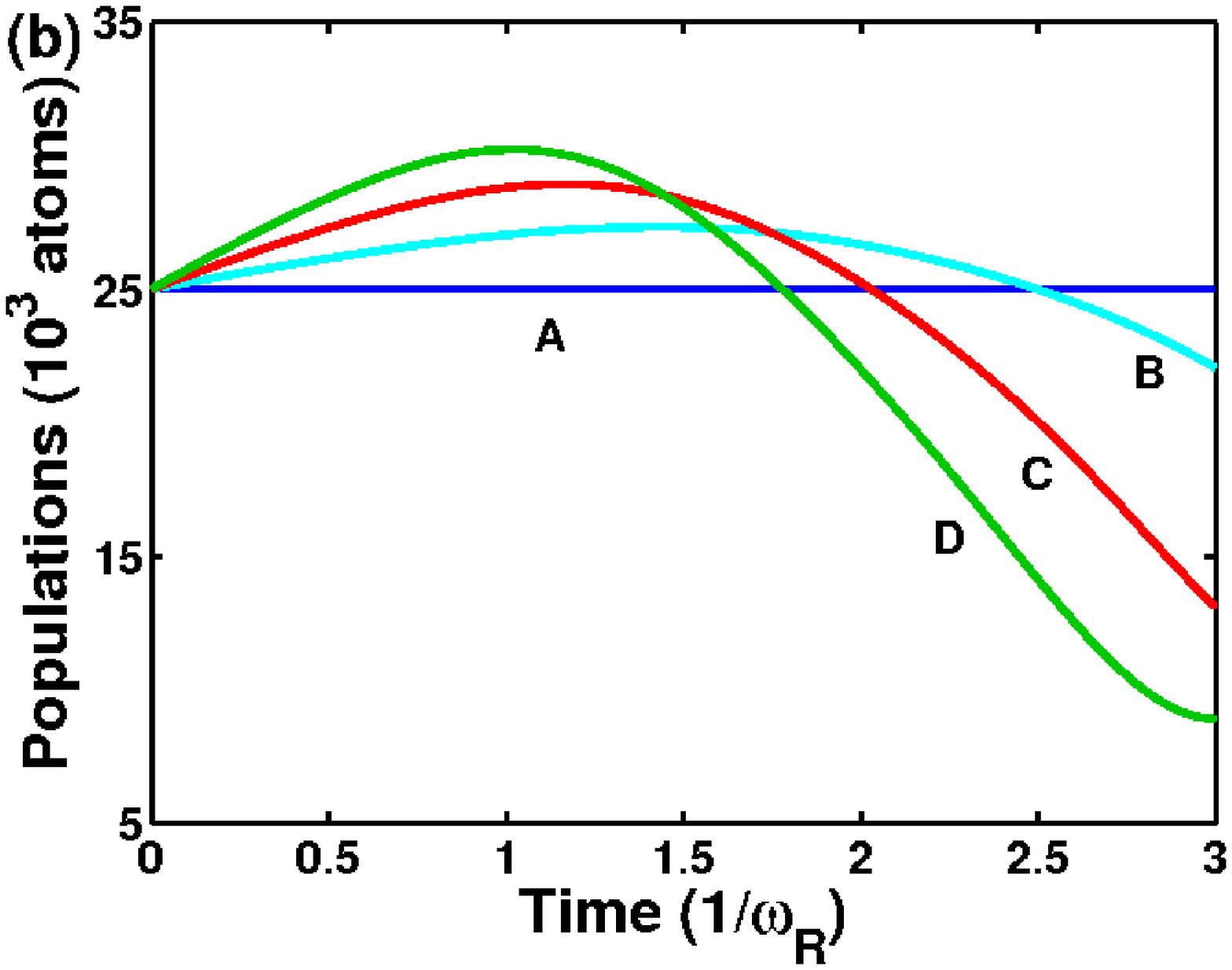}
\caption{\label{fig3}(Color online) (a) Populations versus time plot for a 
system with a phase difference of $2\pi$ around the loop. For $t<0.5/\omega_R$ there is a steady flux and the populations stay constant, for $t>0.5/\omega_R$ the link between two wells is cut 
and by inertia the condensate tends to flow toward the last well before the cut, finally for $t>2/\omega_R$ the inertial effect is exhausted and the condensate starts to flow back. The lines A,B,C and D are relatives to the first, second, third and fourth well respectively. 
(b) System with a persistent current in presence of a coupling constant different from the others. The population of one well is plotted for several values of this coupling constant ($K_{A}=\tilde K, K_{B}=1.2 \tilde K, K_{C}=1.4 \tilde K,  K_{D}=1.6\tilde K$). In these simulations $N_T=10^5$ and $\Lambda=100$.}
\end{center}
\end{figure}

In order to probe what is happening we simulate an abrupt cut of the loop at one point (by setting one coupling constant to zero for $t>\tau$) and observe at the results, shown in Fig. \ref{fig3}. If effectively there was a steady flow we would expect it to continue by inertia so that the last well before the cut would be filled with atoms while the others would be depleted;
this is exactly what emerges from Fig. \ref{fig3} (a).
In experiments it is often quite difficult to control the exact value
of the coupling constants and also to assure that all of them have 
the same value. A coupling constant different from the 
others will behave as a bottleneck, thus creating an oscillation of populations superimposed on the steady flux  (Fig. \ref{fig3} (b)). In order to perform any experiment it would then be necessary to assure that these oscillations do not hide the current we are interested in. In any case this turns out not to be a major problem as both the period and the amplitude of these oscillations increase with $\Lambda$. It is then a priori possible, knowing that we have a given degree of incertitude in the coupling constants, to tune the value of $UN_T$ in order to have enough time to
perform our measurements before large oscillations develop.

\subsection{Self-confinement}

As in the two-well case (\cite{smer97}) the nonlinearity of the GPE leads to the possibility of self-confined states.
 
As a condition for a self-confined state to arise we can take that
in the time in which the phase difference between two wells varies
between $0$ and $\pi$ their population difference doesn't change sign. 
Using Eq.(\ref{realsystem}) we can derive, by means of very rough approximations, 
an approximate condition for this to happens:

\begin{equation} 
\label{condition}
\frac{2}{3n}\tan(\frac{-3\sqrt3}{4\Lambda n})+1>0.
\end{equation}
  
\begin{figure}
\begin{center}
\includegraphics[width=4.2cm,height=3.2cm]{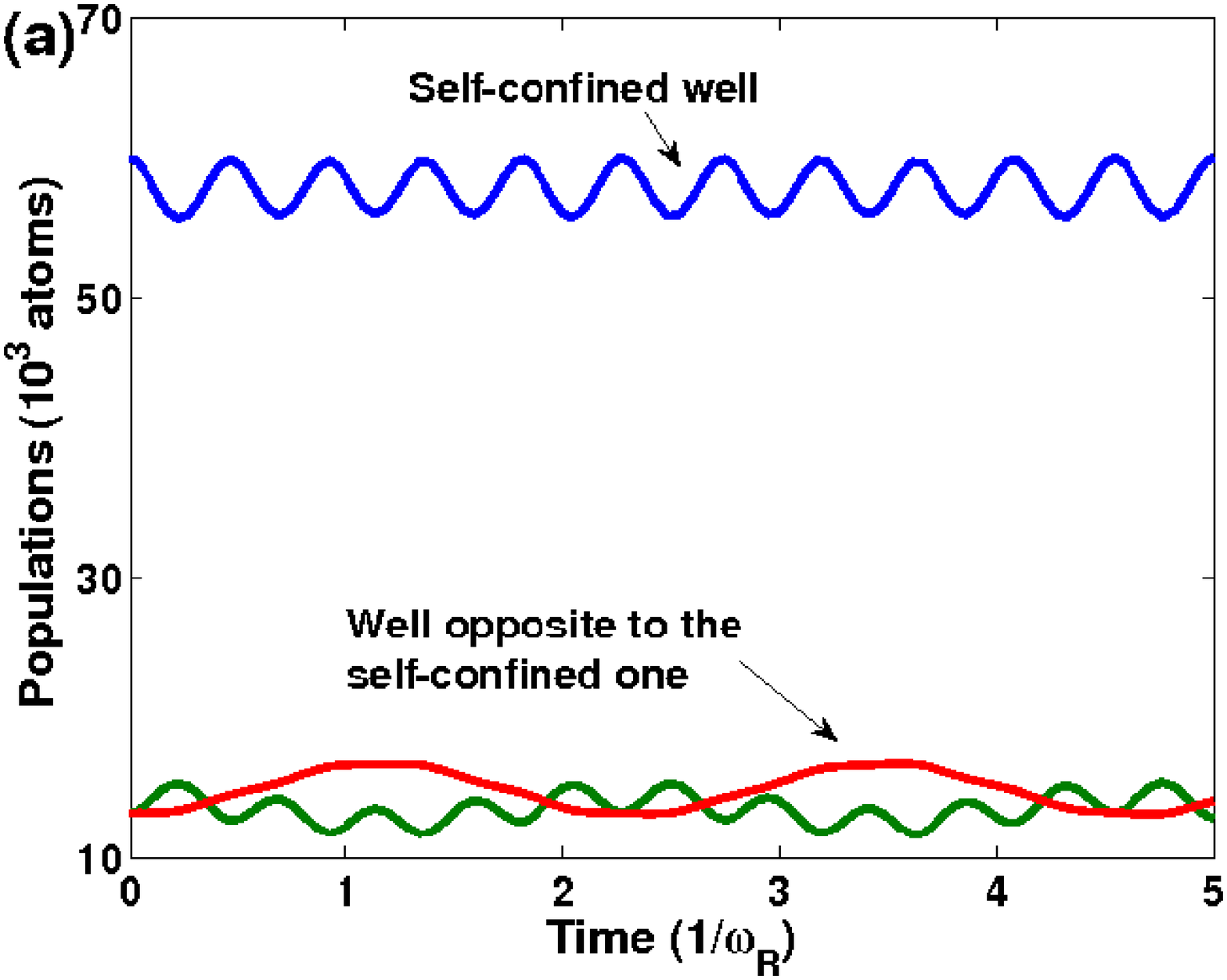}
\includegraphics[width=4.2cm,height=3.2cm]{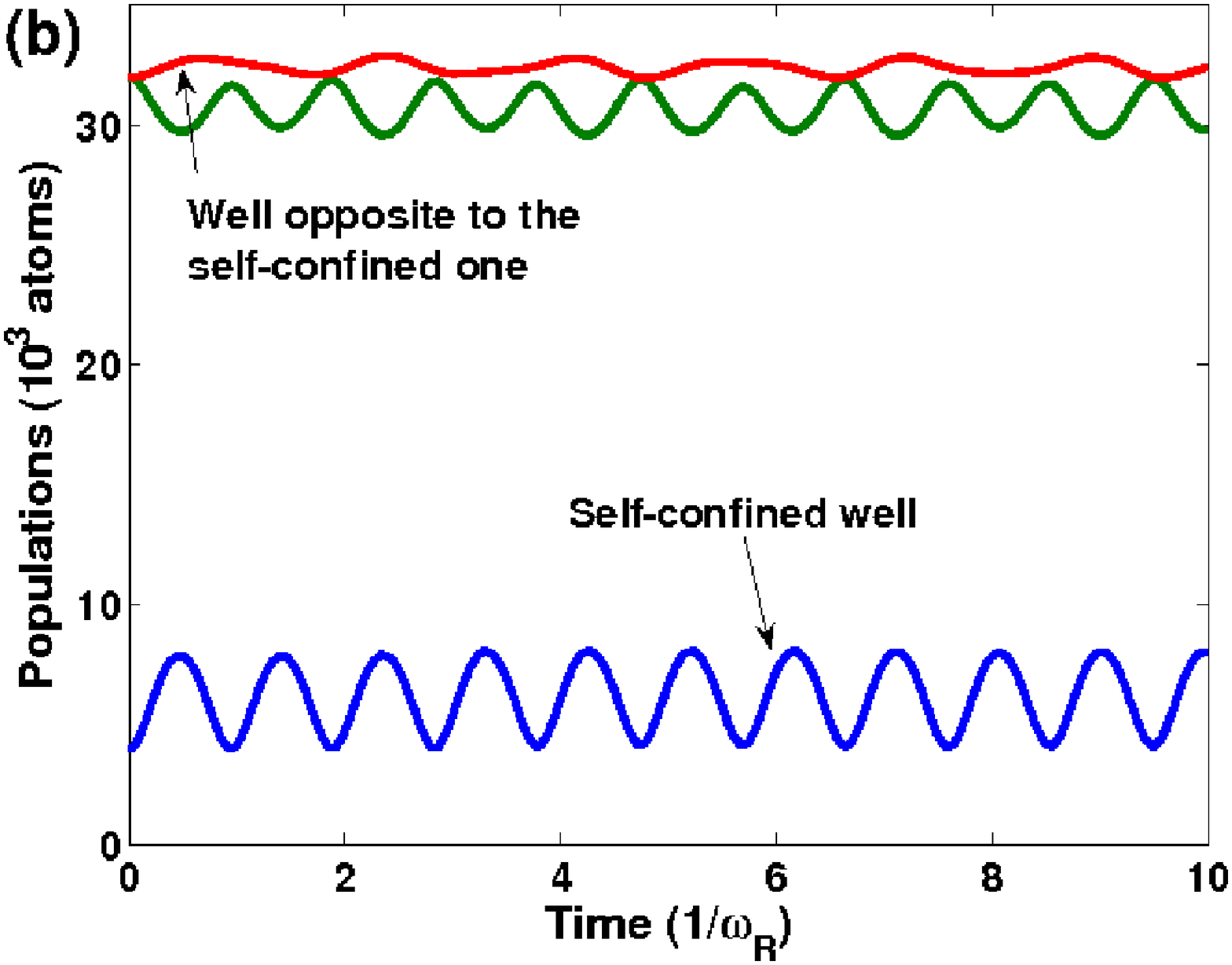}
\caption{\label{fig4}(Color online) Populations versus time plot of (a) the self-confinement
and (b) self-depletion regimes for $\Lambda=100$ and $N_T=10^5$. Due to the symmetry of our system two populations have the same value and so only three lines are visible.}
\end{center}
\end{figure}

Where $n$ is the normalized population imbalance between the well in which we are considering the self-confinement and the population of an adjacent well, supposed to be $\frac{N_{T}-N}{3}$. 
We can check that this condition is satisfied both for positive and negative values of $n$.
The first case corresponds to the usual quantum self-confinement, the second one corresponds to a state of self-depletion, in which one well almost empty at the beginning remains almost empty.
Numerical simulations show that condition stated in Eq.(\ref{condition}) 
works effectively quite well. For $\Lambda = 100$ and $N_T=10^5$ it gives us a minimal number of atoms in one well for being in the self-confined regime of $31750$ while the simulated value is $35000$ and a maximal number of atoms in the considered well for being in the self-depleted regime of $18250$ where the simulated value is $15000$.
It is important to remark that in the second case
the populations of the other three wells are smaller than the lower limit
for the quantum self-confinement, so effectively we are considering a new effect, not simply a self-confinement of the three full wells.
We can have an insight into this phenomenon if we interpret the usual 
quantum self-confinement of the two-well system as a self-confinement for
one well and a self-depletion for the others. In the two-well system the 
conditions for having them coincide (obviously if one well is empty the other has to be full) but in the four wells case they don not coincide anymore and we can thus experimentally distinguish them (Fig. \ref{fig4}). In both cases we will have to face the standard dissipation phenomena we encounter in the usual quantum self-confined regime (\cite{kohl03,zapa03}) but also in this case the time-scale we are interested in to observe the phenomenon are substantially smaller when compared with the typical dissipation time.

\subsection{Full condensate spinning}
\begin{figure}
\begin{center}
\includegraphics[width=8.6cm,height=5.8cm]{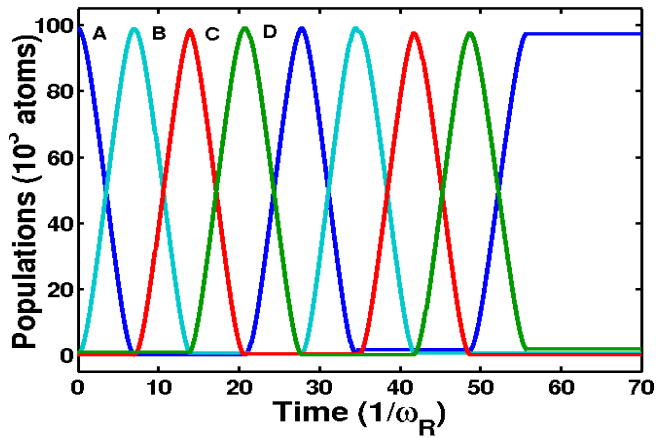}
\caption{\label{fig5}(Color online) The condensate is moved around the loop by making the different wells to switch between the normal and the self-confined regime. The lines A,B,C and D are the populations respectively of the first, the second the third and the fourth well. After two full turn the modulation is stopped and the spin ceases. In this simulation $N_T=10^5$ and $\Lambda=100$.}
\end{center}
\end{figure}

Exploiting the self-confining nonlinearity we can make the 
entire condensate move all the way around the loop. 
Let us suppose that the condensate is at the beginning at one well.  
Increasing the coupling constant toward an adjacent well, enough in order to
make the well exit the self-confined regime, causes a flux 
toward this second well. By lowering the coupling constant before 
the flux changes sign, we find ourselves with the second well in the 
self-confined regime. Repeating this process we can 
make the condensate move around the loop  (Fig. \ref{fig5}) in a sort of quantum version of the conveyor belt mechanism (\cite{hans00}). The number of atoms in the depleted wells has to be quite small (of the order of few percents of $N_T$) in order to avoid having significant spurious population's oscillations that could bring the system off resonance after very few turns.
The process has revealed itself to be almost independent from the initial phases of the four condensates.
This spinning of the whole BEC is possible because we have no intensity 
resonance, as we had before, and so the resonance frequency remains the same. In any case, because of the long period implicated (of the order of some centains of $1/\omega_R$), dissipation could play an important role. However, contrary to what happens in high frequency oscillation in one self-confined well, here we have only very low frequency population's fluctuations. Thus the only significant dissipation source should be the spontaneous atom loss that is normally possible to keep relatively small for the time-scale we are interested in. It is not straightforward to calculate the correct modulation frequency analytically but simple numerical integration of Eq.(\ref{firstsystem}) is sufficient to find it accurately enough for any experiment. We may notice however that contrary to what happens in the others two cases, as  soon as the modulation is interrupted, the spinning stops.
 
\subsection{Conclusion}
We have shown that, even though the dynamics of BEC in two-well system is a well studied problem, the extension to four wells allows us to highlight new and interesting aspects. It also gives us the possibility to understand better well known phenomena. We would like to stress that we used the four wells configuration simply because it is the less complex nontrivial loop configuration, but that the phenomena we highlighted have nothing specific to the number four, being all of them trivially expandable to loops made up of a larger number of wells. 
S.D.L. would like to thank Yvan Castin for useful comments and corrections.

\bibliography{manuscript}
\end{document}